\documentclass[a4paper,11pt]{article}
\usepackage{pos}

\title{Diffusive Shock Acceleration of Cosmic Rays -- Quasi-thermal and Non-thermal Particle Distributions}
\ShortTitle{Diffusive Shock Acceleration of Cosmic Rays}

\author{Bojan Arbutina}

\affiliation{Department of Astronomy, Faculty of Mathematics, University of Belgrade,\\
  Sudentski trg 16, 11000 Belgrade, Serbia}

\emailAdd{arbo@matf.bg.ac.rs}

\abstract{A well-known paradigm about the origin of Galactic cosmic rays (CRs) is that these high-energy particles are accelerated in the process of diffusive shock acceleration (DSA) at collisionless shocks (at least up to the so-called "knee"energy of $10^{15}$ eV). Knowing the details of injection of electrons, protons and heavier nuclei into the DSA, their initial and the resulting spectrum, is extremely important in many "practical" applications of the CR astrophysics, e.g. in modelling of the gamma or synchrotron radio emission of astrophysical sources. In this contribution I we will give an overview of the DSA theory and the results of observations and kinetic Particle-In-Cell (PIC) simulations that support the basic theoretical concepts. PIC simulations of quasi-parallel collisionless shocks show that thermal and supra-thermal proton distribution functions at the shock can be represented by a single quasi-thermal distribution - the  $\kappa$-distribution that is commonly observed in out-of-equilibrium space plasmas. Farther downstream, index $\kappa$ increases and the low-energy spectrum tends to Maxwell distribution. On the other hand, higher-energy particles continue through the acceleration process and the non-thermal particle spectrum takes a characteristic power-law form predicted by the linear DSA theory. In the end, I will show what modification of the spectra is expected in the non-linear DSA, when CR back-reaction to the shock is taken into account. }

\FullConference{%

11th International Conference of the Balkan Physical Union (BPU11),\\
  28 August - 1 September 2022\\
  Belgrade, Serbia
}

\begin{document}
\maketitle

\section{Introduction}
As widely known, cosmic rays (CRs) were discovered by Victor Hess in his famous balloon experiment in 1912 (see \cite{Longair2006}), for which he later won the Nobel prize in physics in 1936. CRs are primarily composed of protons, $\alpha$-particles and heavier nuclei, but also electrons. We can actually distinguish between primary and secondary CRs, the latter being produced in Earth's atmosphere by primary particles and observed as air showers --  extensive cascades of ionized particles and photons, which can be many kilometres wide (see e.g. \cite{Longair2011}).
We can also distinguish between CRs at the top of Earth atmosphere and CRs at a source (which can, excluding energetic particles coming from the Sun, be Galactic or extragalactic). The observed all particle distribution is largely isotropic and to a very good approximation a power-law with the slope $\mathnormal{\Gamma} \approx 2.7$ up to the energy $\sim 10^{15}$ eV -- the so-called "knee" in the spectrum (Fig. 1). At lower energies, the spectrum is affected by Solar modulation. Above the "knee" the spectrum is steeper, ending in the so-called "ankle" and "toe" region at about $\sim 10^{19} - 10^{20}$ eV. The dominant paradigm about the origin of CRs is that they are majorly produced by sources within our Galaxy, such as supernova remnants (SNRs) that might be able to produce CR particles even up to the $\sim 10^{17}$ eV \cite{Morlino2016}. It is believed that higher-energy and ultra high-energy CRs are of extragalactic origin.

CRs produced by Galactic sources are affected by magnetic field and diffuse through the Galaxy so the observed CRs spectrum is expected to be different than the spectrum at the source. In this contribution, we are primarily interested in the latter and we will  focus on strong shocks of SNRs as the sites of particle acceleration through the so-called first order Fermi acceleration.  Originally, a mechanism for particle acceleration at the source (interstellar medium (ISM) clouds acting as magnetic mirrors) was proposed by Fermi \cite{Fermi1949} and this is now called type II or the second order Fermi acceleration, because the relative gain in particle energy is $\Delta E/E \sim (V/c)^2$, where $V$ is the relative velocity of clouds that reflect particles. In the more efficient type I or the first order Fermi acceleration $\Delta E/E \sim V/c$. A
modern version of the first order Fermi acceleration -- diffuse shock acceleration (DSA) theory, in which $V$ is actually the shock velocity, was developed
independently by Axford et al., Krymsky, Bell,
and Blandford and Ostriker \cite{Axford1977, Krymsky1977, Bell1978, BO1978}. There are two main approaches to
the problem: macroscopic and
microscopic developed by Bell \cite{Bell1978}. In the next section we will give a brief overview of the basic concept of DSA and write down the resulting non-thermal particle spectrum.

\section{Non-thermal particle distribution}

DSA is in its essence a non-thermal process. In the macroscopic approach, one starts
from collisionless plasma kinetic equation to find the momentum distribution function $f(p)$ immediately ahead (upstream) and behind (downstream) the shock front. Since this function should be invariant across the shock, from matching conditions one finds the power-law spectrum $f(p) \propto p^{-\mathnormal{\Gamma} -2}$ where the slope is related to the shock compression $R$ as $\mathnormal{\Gamma} =(R+2)/(R-1)$ (see e.g. \cite{Drury1983, BE1987}). The microscopic approach is more intuitive since it tries to explain what is happening with individual particles. The basic idea behind it is that if (seed) particles are energetic enough, they can cross and re-cross the shock front unaffected, unlike the thermal particles that are advected and heated downstream by the shock. These seed particles gyrating along (nearly) parallel magnetic field $B_0$ are being scattered by turbulances/Alfven waves behind/ahead of the shock which act as scattering centers in the downstream/upstream plasma. If we make Lorentz transformations from upstream to downstream plasma frame, we will see that in each crossing and re-crossing of the shock a particle has gained a small amount of energy i.e. momentum $\Delta p/p \propto V/c$. If there are many such crossings, a particle will ultimately become ultra-relativistic. What is happening in parallel at strong shocks is magnetic field amplification due to streaming instability \cite{Bell2004, AB2005}, in addition to plain shock compression (of normal field component) which cannot explain the observed magnetic field strengths.

\begin{figure}[t]
\center{
  \includegraphics[bb=0 0 789 388,angle=0,width=\textwidth,keepaspectratio]{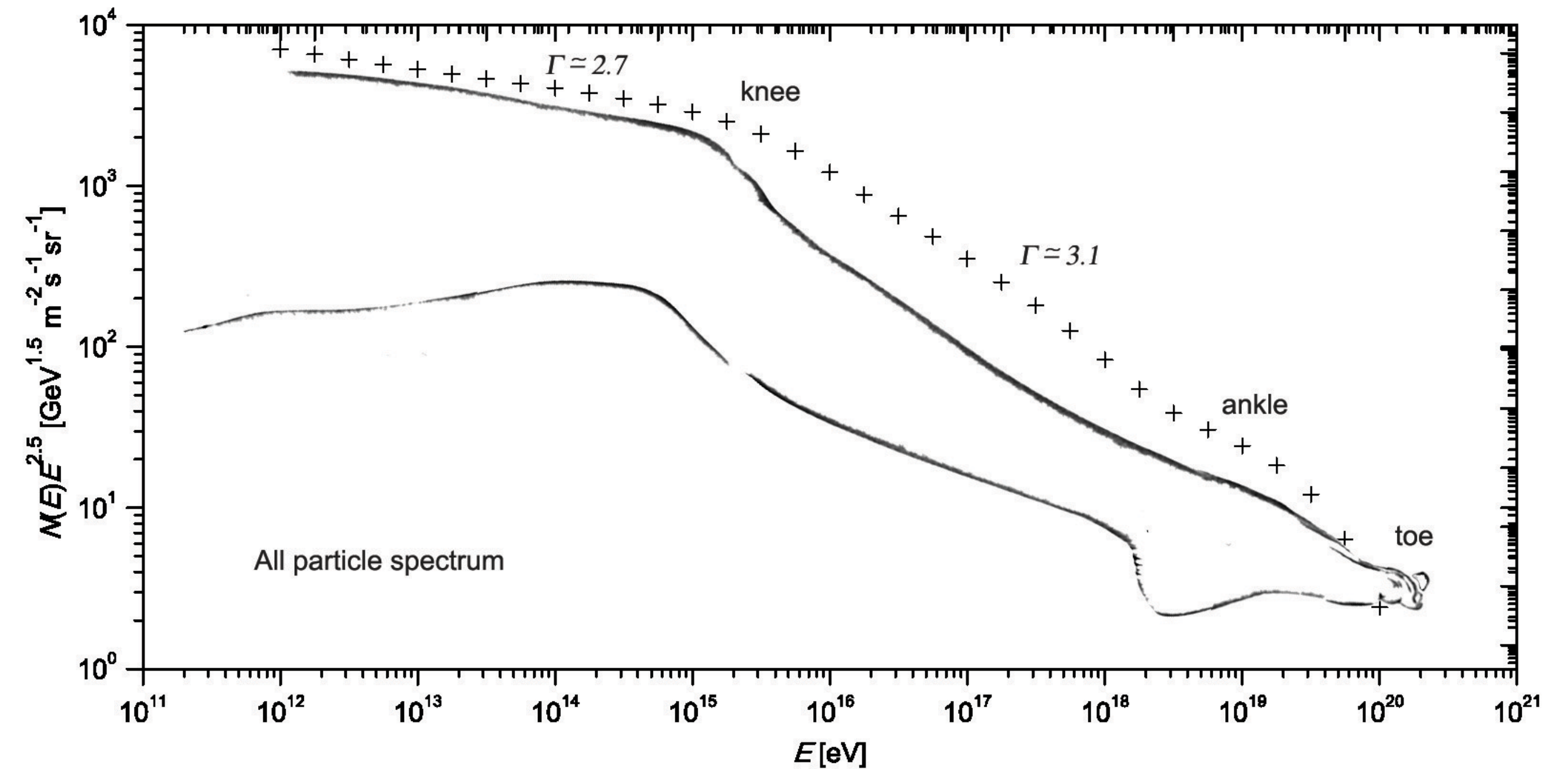}
  \caption{Observed cosmic rays energy spectrum \cite{Arbutina2017}. Data are from H4a model \cite{Gaisser2012}.}}
\end{figure}

To find the particle spectrum, we will follow the derivation from \cite{AZ2021b}. We will not assume that the particles are ultra-relativistic ($v \neq c$), so for the momentum gain we have
\begin{equation}
\mathcal{G} =  \frac{\Delta p}{p} \approx \frac{4}{3}\frac{V_1-V_2}{v} = \frac{4(R-1)}{3}\frac{V_2}{v} ,
\label{eq1}
\end{equation}
where $V_1 = -V$ and $V_2=V_1/R$ are respectively upstream and downstream plasma velocity, as observed in the shock frame (upstream plasma is assumed to be at rest in the laboratory frame). Bell \cite{Bell1978} argued that the probability of a particle to be advected downstream is $\mathcal{P} = \frac{4V_2}{v}$, and consequently, the probability to escape upstream and stay in the process of acceleration is $\mathcal{P}_B = 1 -  \mathcal{P}$. In \cite{AZ2021b}, the authors generalized Bell's probability to (see also \cite{Vietri2008, Blasi2012}):
\begin{equation}
\label{eq5} \mathcal{P}_B = \Bigg( \frac{1-\frac{V_2}{v}}{1+\frac{V_2}{v}}\Bigg) ^2 \approx 1- \frac{4V_2}{v},
\end{equation}
assuming non-relativistic shocks.

Probability $\mathcal{P}$ is related to the cumulative number change
\begin{equation}
\frac{\Delta N}{N} =  -\mathcal{P} = \mathcal{P}_B -1 = \frac{-\frac{4V_2}{v}}{(1+V_2/v)^2},
\end{equation}
which combined with the momentum gain, under transformations
$
\frac{\Delta N}{N}  \rightarrow \frac{dN}{N} = -\mathcal{P}
$ and $
\frac{\Delta p}{p}  \rightarrow \frac{dp}{p} = \mathcal{G}
$
give
\begin{equation}
\frac{d\ln N}{d\ln p} = -\frac{\mathcal{P}}{\mathcal{G}}, \ \ \ f= - \frac{1}{4\pi p^2}\frac{dN}{dp} .
\end{equation}
From the last equation, one can finally obtain
\begin{equation}
\label{sol}
f(p) = \frac{3N_\mathrm{CR}}{4\pi (R-1) p_{0}^3}  \Big(1+\frac{V_2}{v_{0}}\Big)^{\frac{3}{R-1}} \Big(\frac{p}{p_{0}}\Big)^{-\frac{3R}{R-1}} \Big(1 +  \frac{V_2}{v}\Big)^{-\frac{2R+1}{R-1}}   e^{\frac{3  V_2}{R-1}(\frac{1}{v_{0}+V_2} - \frac{1}{v +V_2})},
\end{equation}
where $N_{CR}$ is the total number of CRs, and $p_{0}$ is some injection momentum. For $p \gg m V_2$ Eq. (\ref{sol}) gives the same power-law dependence as the microscopic approach $f(p) \propto (p/p_{0})^{-3R/(R-1)} \propto p^{-4}$ i.e. $\mathnormal{\Gamma} =2$ for the standard shock compression $R=4$. The non-thermal CR proton and electron distributions are shown in Fig. 2, assuming $N_\mathrm{p} = N_\mathrm{e}$ and equal injection energies (as we shall see later, this implies nearly equal temperatures $T_\mathrm{p} \approx T_\mathrm{e}$). The high-energy power-law distribution is well supported (at least for electrons) by radio observations of astrophysical sources, SNRs in particular, which show power-law frequency spectra $\propto \nu^{-\alpha}$ where spectral index $\alpha = (\mathnormal{\Gamma} -1)/2$ has a mean value around 0.5 (again $\mathnormal{\Gamma} \approx 2$). Distributions at lower momentum given by Eq. (\ref{sol}) and plotted in Fig. 2 are uncertain, since here we enter the supra-thermal particles domain.

\begin{figure}[t]
\center{
  \includegraphics[bb=50 0 770 600,angle=0,width=0.9\textwidth,keepaspectratio]{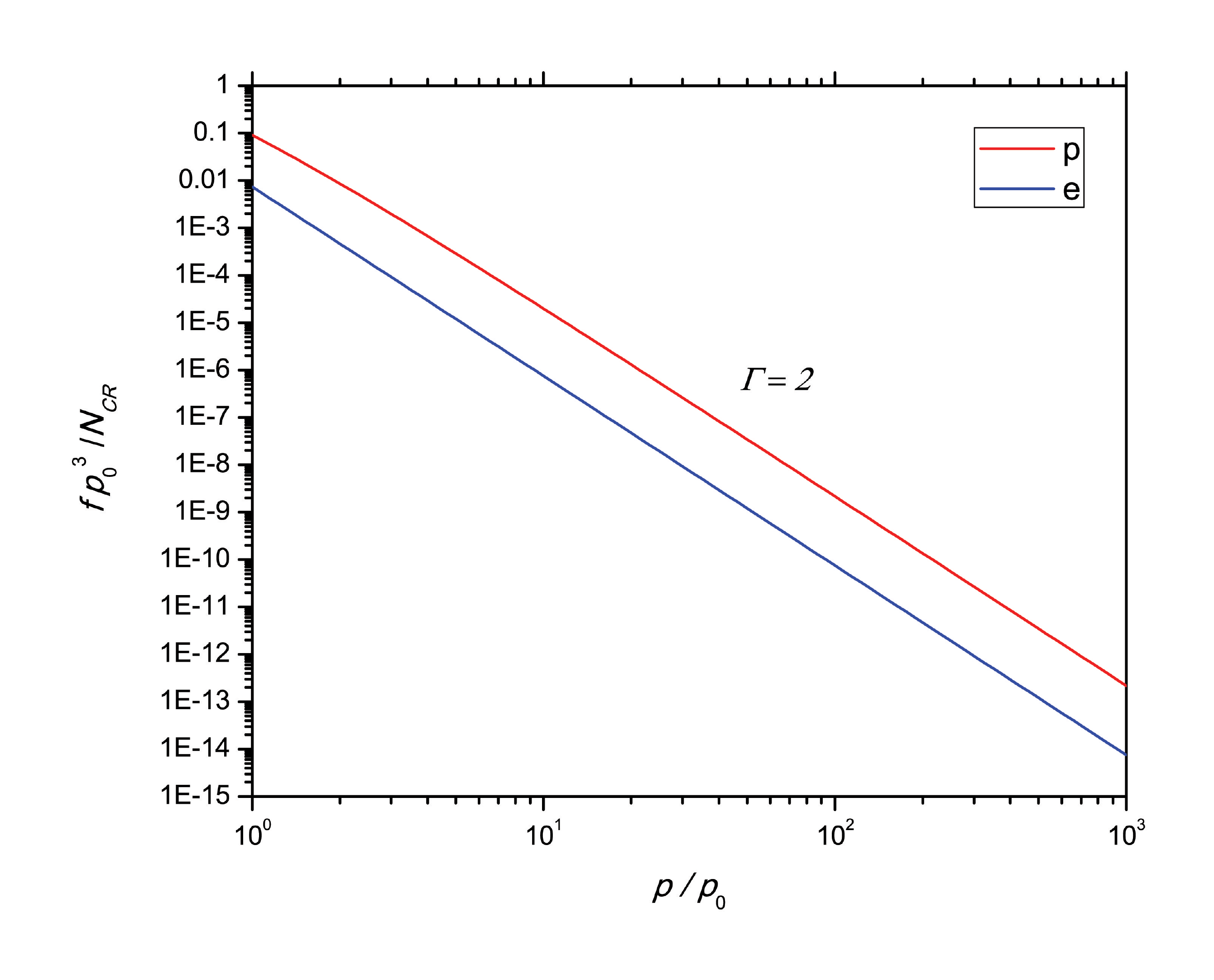}
  \vspace{-0.25cm}
  \caption{The non-thermal CR proton and electron distributions with $\mathnormal{\Gamma} = 2$, obtained by assuming $N_\mathrm{CR}=N_\mathrm{p} = N_\mathrm{e}$ and equal injection energies $E_0 = \frac{p_0^2}{2 m_\mathrm{p}}\sim \frac{1}{2} m_\mathrm{p} V^2$.}
  }
\end{figure}

\section{Supra-thermal particles}

Since observational data about particle acceleration is limited, important information  and valuable insight in this process are obtained through kinetic particle-in-cell (PIC)  or hybrid simulations of collisionless shocks that serve as a sort of astrophysical laboratories (see e.g. \cite{CS2014a, CS2014b, CS2014c}). They are particularly important for understanding the injection mechanisms. The acceleration of CR protons is generally better understood, because if the shock thickness is of the order of gyroradius of thermal protons, a small number of slightly more energetic protons should be able to move across the shock and thus be injected into DSA. The acceleration of electrons is generally less understood, since due to their much lower mass, electrons have significantly smaller gyroradii than protons and are thus more bound to shock front. Nevertheless, they seem to be pre-accelarated through shock-drift acceleration (SDA) or by a combination of SDA and DSA, until they reach injection momentum of protons, enter full DSA cycles and continue to behave in a similar fashion as ions, ultimately becoming ultra-relativistic and following a characteristic non-thermal power-law test-particle DSA distribution (see \cite{Park2015, Guo2015, AZ2021a}).

\begin{figure}[t]
\center{
  \includegraphics[bb=0 0 1800 1800,angle=0,width=0.9\textwidth,keepaspectratio]{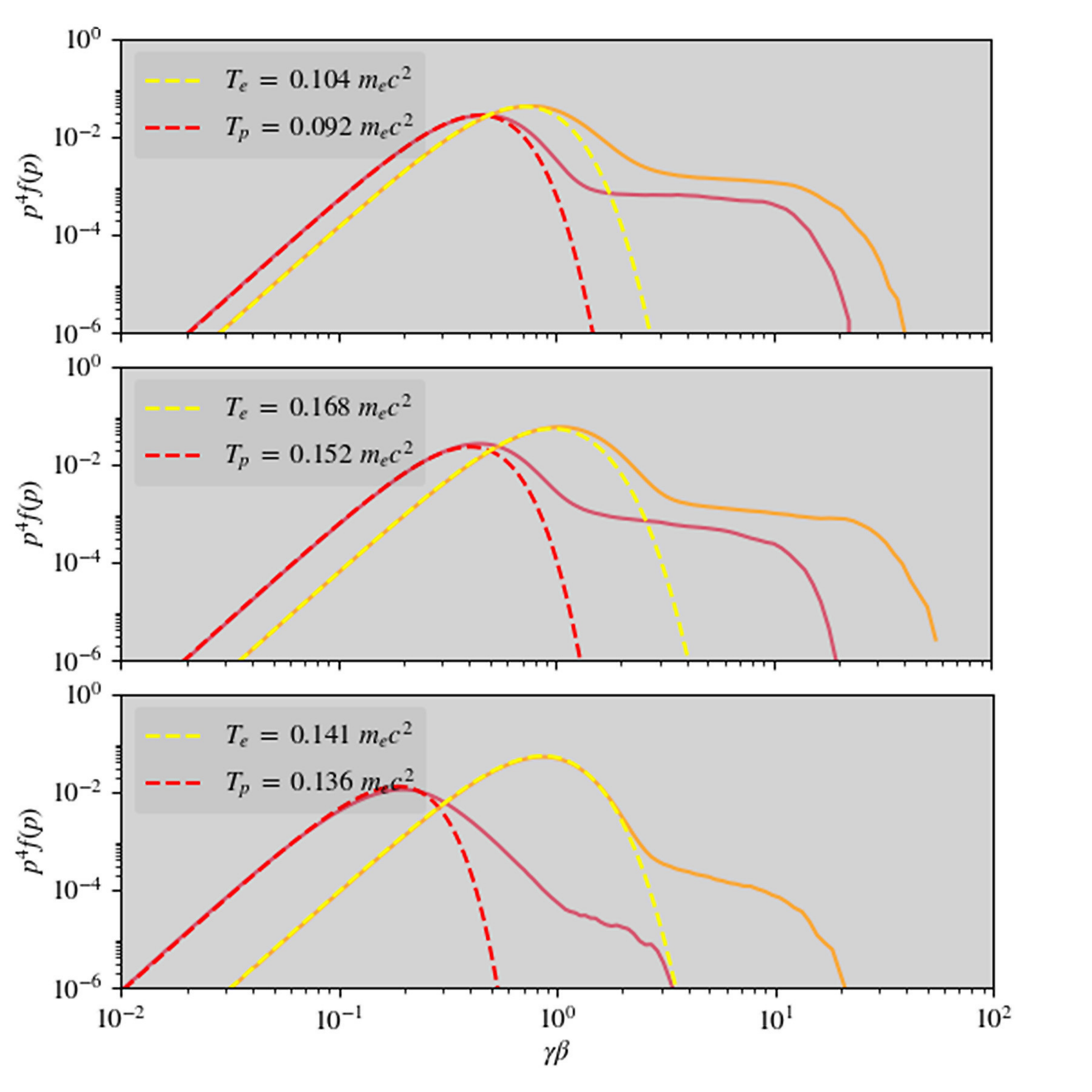}
  \vspace{-0.1cm}
  \caption{Proton and electron spectra in simulation runs 1-3 (with mass ratio $m_\mathrm{p}/m_\mathrm{e}$ = 2, 4, 16, from top to bottom, respectively) from \cite{AZ2019}. The simulation parameters are: magnetization $\sigma = B_0^2/(4\pi \gamma _0 m_\mathrm{p} n_\mathrm{p} c^2)) = 2.5\times 10^{-3},\ 2.5\times 10^{-3},\ 0.6\times 10^{-3}$, shock velocity $V/c = 0.67,\ 0.67,\ 0.33$ and Alfvenic Mach number $M_A = 13$, where $B_0$ is ambient magnetic field, $n_\mathrm{p}$ proton number density, $\gamma _0 = 1/\sqrt{1-V^2/c^2}$.}
  }
\end{figure}

In Fig 3. we give spectra obtained in PIC simulations of quasi-parallel collisionless shocks with different proton-to-electron mass ratios from \cite{AZ2019}. We see that proton and electron spectra have similar shapes that at low energies deviate from the Maxwellian (given by dashed lines). A particularly visible deviation is in the supra-thermal domain, between thermal and non-thermal energies i.e momenta.
In \cite{Caprioli2015} the authors accomplished to explain this non-thermal distribution by the so-called minimal model. The idea is that while most of the ions will be advected and thermalized downstream, some can gain extra energy in SDA  or micro-DSA \cite{AZ2019} cycles becoming supra-thermal. Those protons that continue DSA cycles will become non-thermal. The complete distribution function can thus be represented as a combination of thermal (Maxwelian), supra-thermal and non-thermal (power-law) parts.

To find the supra-thermal distribution, Caprioli et al. \cite{Caprioli2015} assumed a constant  probability for a particle to be advected $\mathcal{P} = 0.75$, meaning that roughly 75\% of particles would be thermalized, while the remaining 25\% would become supra-thermal/non-thermal. This can be understood if there is a modified probability for a particle to cross to the upstream i.e. stay in the cycles, $\mathcal{P}_A \cdot \mathcal{P}_B$, where $\mathcal{P}_B$ is Bell's probability and $\mathcal{P}_A$ a probability for a particle to pass through the shock of some finite thickness, not being reflected back downstream. If again we use the approach from \cite{AZ2021b} with momentum gain from Eq. (\ref{eq1}) and assuming the particles being non-relativistic, $p = mv$, we get for the supra-thermal distribution
\begin{equation}
f(p) = \frac{3N_\mathrm{ST} \mathcal{P}}{16\pi (R-1) m V_2 p^2}  {e^{\frac{3 \mathcal{P}}{4(R-1)m V_2}(p_\mathrm{min} - p)}} ,
\end{equation}
where $N_\mathrm{ST}$ is the total number of suprathermal particles, and $p_\mathrm{min}$ is the minimum momentum at which particles enter SDA.


Instead of describing supra-thermal and non-thermal particle distributions through the same formalism, one could also try to describe thermal and suprathermal particle distribution with one quasi-thermal distribution -- the $\kappa$-distribution \cite{AZ2020, AZ2021b}. Such non-equlibrium distributions are common to space plasmas, with index $\kappa$ being a free parameter which is a kind of a measure of non-equilibrium \cite{LM2011, Livadiotis2017}. When $\kappa \to \infty$, the plasma reaches the thermodynamic equilibrium and the distribution becomes Maxwellian.

Relativistic generalization of $\kappa$-distribution can be written as
\begin{equation}
\frac{dN}{dp} = 4\pi p^2 f = \frac{N_0 C p^2}{m^3 c^2} \frac{1}{\Bigg[ 1+ \frac{\sqrt{1+\frac{p^2}{m^2c^2}}-1}{\kappa \Theta} \Bigg]^{\kappa +1}},
\end{equation}
where constant $C$ is found from normalization $\int _0 ^\infty 4\pi p^2 f dp=N_0$:
\begin{equation}
C^{-1} = \frac{(\pi \kappa ^2 \Theta ^2)^{3/2} (\kappa +1)\Gamma (\kappa -2) _2F_1 (-\frac{3}{2},\kappa  -2,\kappa  +\frac{1}{2},1-\frac{2}{\kappa \Theta})}{4\pi \Gamma (\kappa +\frac{1}{2})},
\end{equation}
$\Gamma (x)$ being the Gamma-function, $_2F_1 (a,b,c,x)$ is Gauss hypergeometric function (see \cite{AS1972}) and $\Theta = kT_\kappa/(mc^2)$, $T_\kappa$ not being the usual thermodynamic temperature if plasma is out of equilibrium.

In the limit $\kappa  \to \infty$, this distribution becomes the Maxwell-J\"utter distribution (see \cite{Synge1957})
\begin{equation}
\frac{dN}{dp} = \frac{N_0 p^2}{m^3 c^2 \Theta K_2(1/\Theta)} e^{-\frac{\sqrt{1+\frac{p^2}{m^2c^2}}}{\Theta}},
\end{equation}
where $T$ is now thermodynamic temperature and $K_2(x)$ the modified Bessel function of the second order.
On the other hand, if $\Theta \ll 1$ ($p \ll mc$) we can obtain standard non-relativistic $\kappa$-distribution
\begin{equation}
\frac{dN}{dp} = \frac{N_0 4 \pi p^2}{(\pi \kappa p_\kappa ^2)^{3/2}} \frac{\Gamma(\kappa +1)}{\Gamma(\kappa -\frac{1}{2})} \frac{1}{\Big[ 1+ \frac{p^2}{\kappa p_{\kappa} ^2} \Big]^{\kappa +1}},\ \ \ p_\kappa ^2 = 2 mkT_\kappa
\end{equation}
which, as we said, for $\kappa  \to \infty$ tends to Maxwell distribution
\begin{equation}
\frac{dN}{dp} =  \frac{4\pi p^2 N_0}{(2\pi m kT)^{3/2}} e^{-\frac{p^2}{2mkT}}.
\end{equation}

For higher momenta, $\kappa$-distribution is actually a power-law. Note, however, the different power-law dependence of non-relativistic  and ultra-relativistic $\kappa$-distribution when $p \to \infty$
\begin{equation}
 \frac{dN}{dp}  \propto  \Bigg\{ \begin{array}{@{\extracolsep{-1.0mm}}ll @{}}
 \ p^{-2\kappa}, & \ \ \ p\ll mc \\
 \ p^{1-\kappa},  & \ \ \ p \gg mc .
 \end{array}
 \label{m17}
\end{equation}
This is relevant for ultra-relativistic shocks when one component (namely, electrons) can be highly relativistic (e.g. jets of active galactic nuclei). In the proceeding section, as it was in the major part of the manuscript, we shall, nevertheless, deal only with non-relativistic shocks, characteristic for SNRs.

\section{Non-linear diffusive shock acceleration}

Test-particle approach to DSA assumes that the energy density or pressure of CRs is negligible, so that CR presence does not modify (Rankine-Hugoniot) jump conditions. When this is not the case, we are talking about CR back-reaction and non-linear DSA (see e.g. \cite{Drury1983, MD2001, Blasi2002a, Blasi2002b, AB2005}). The back-reaction of CRs  affects the shock in such a way that high-energy particles ahead of the shock induce shock precursor with density, pressure and velocity gradients. The discontinuity is, nevertheless, still present as the so-called subshock with compression $R_\mathrm{sub}$. The total compression of a modified shock is larger $R_\mathrm{tot} > R_\mathrm{sub}$.

On the other hand, the shock itself modifies CR particles, producing the concave-up spectrum. This is because the lower-energy CRs will only experience the jump at the subshock and have steeper spectrum, while the higher-energy particles will diffuse farther upstream in the precursor and experience larger compression, consequently having a flatter spectrum.

For modelling proton and electron spectra we shall use Blasi's semi-analytical model whose details can be found in \cite{Blasi2002a,Blasi2002b} (see also \cite{Blasi2004, Blasi2005, Blasi2007, AB2005, Ferrand2010, Pavlovic2018, Urosevic2019, AZ2021a}). Blasi's model implies solving coupled equations (obtained from diffusion-advection equations and mass and momentum conservation):
\begin{equation}
\frac{1}{3}\Big( \frac{1}{R_{\mathrm{tot}}} - U_p\Big) p \frac{d f_\mathrm{p}}{d p} - \Big( U_p + \frac{1}{3} p \frac{d U_p}{d p} \Big) = 0 ,
\end{equation}
\begin{equation}
\frac{1}{3}\Big( \frac{1}{R_{\mathrm{tot}}} - U_p\Big) p \frac{d f_\mathrm{e}}{d p} - \Big( U_p + \frac{1}{3} p \frac{d U_p}{d p} \Big) = 0 ,
\end{equation}
\begin{eqnarray}
\frac{d U_p}{d p} \Bigg( 1 - \frac{U_p^{-(\gamma +1)}}{M_0^2} \Big( 2 + \zeta (\gamma -1) \frac{M_0^2}{M_\mathrm{A}} \Big) - \frac{1 -\zeta}{8 M_\mathrm{A}} \frac{U_p^2 +3}{U_p^{5/2}} \Bigg) \nonumber
\\
=  \frac{p^4 f_\mathrm{p}}{\sqrt{1 + p^2}} + \frac{p^4 f_\mathrm{e}}{\sqrt{(m_\mathrm{e}/m_\mathrm{p})^2 + p^2}},
\end{eqnarray}
where $p$ is in units $m_\mathrm{p} c$, distribution functions $f$ are given in some arbitrary units, $M_0$ is Mach number, $M_A$ Alfvenic Mach number, $\zeta$ Alfven-heating parameter, $\gamma$ adiabatic index and $ U_p = u_p/V$ is dimensionless average velocity in the precursor.
The setup is similar as in \cite{AZ2021a} and assumes constant electron heating ahead of the subshock $
T_{2,\mathrm{p}} = T'_{2,\mathrm{p}}  - \Delta E, \, T_{2,\mathrm{e}} = T'_{2,\mathrm{e}}  + \Delta E
$ where  $\Delta E \approx 0.3$ keV  \cite{Ghavamian2007, Ghavamian2013} and temperatures $T'_2$ are obtained from jump conditions. In any case, the injection parameter $\xi = p_\mathrm{inj}/p_\mathrm{th}$ must be given ($p_\mathrm{th} = \sqrt{2 m_\mathrm{p,e} k T_\mathrm{p,e}}$), and it is assumed that its value is the same for protons and electrons. By matching non-thermal power-law to Maxwell distribution Blasi et al. \cite{Blasi2005} related injection parameter $\xi$ to injection efficiency $\eta=N_{CR}/n$, where $n$ is particle number density,
\begin{equation}
\eta = \frac{4}{3\sqrt{\pi}} (R_{\mathrm{\mathrm{sub}}}-1) \xi ^3 e^{-\xi ^2}.
\end{equation}

\begin{figure}[t]
\center{
  \includegraphics[bb=0 0 768 626,angle=0,width=0.85\textwidth,keepaspectratio]{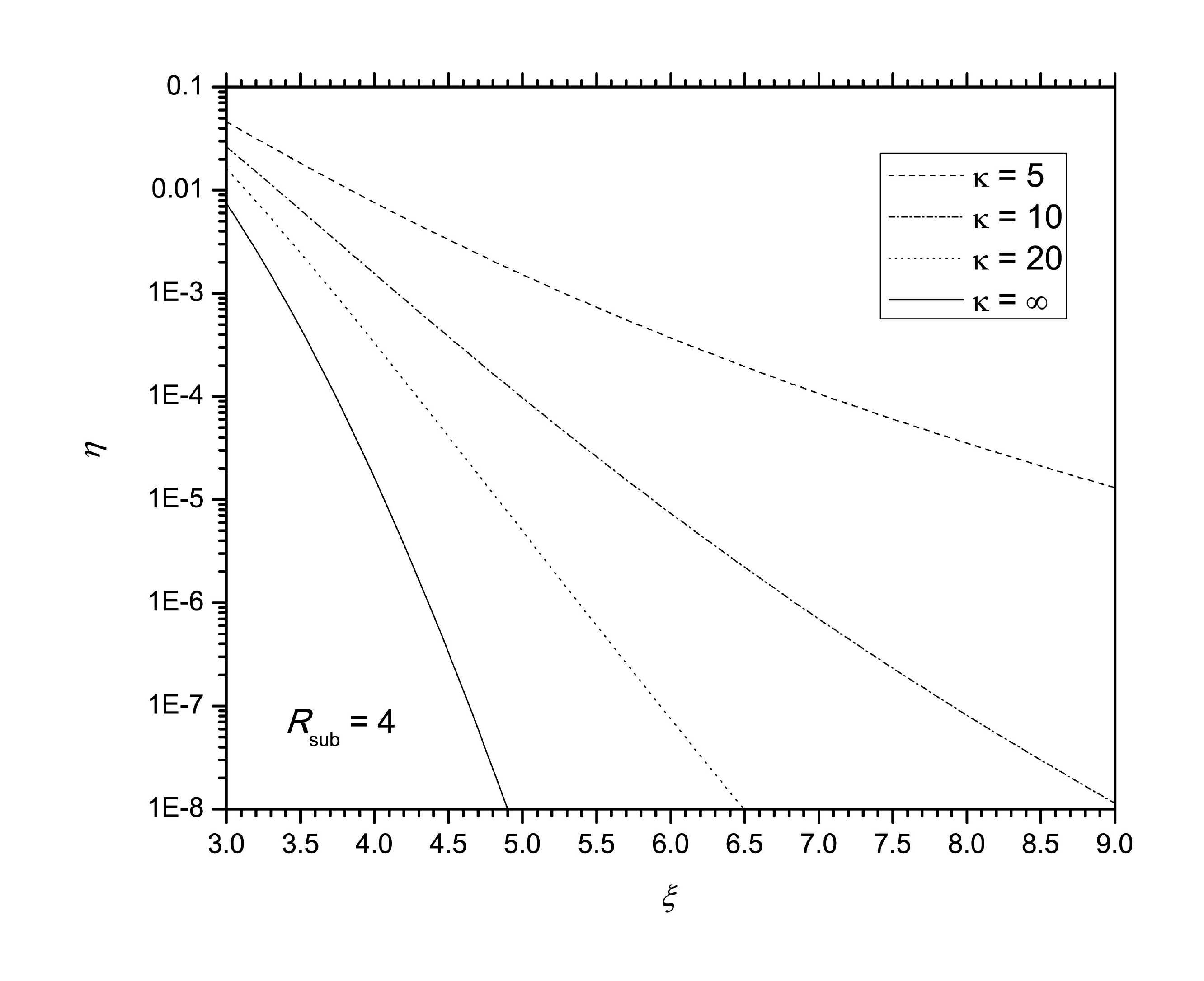}
  \vspace{-0.75cm}
  \caption{Acceleration efficiency as the function of injection parameter for the Maxwellian and $\kappa$ = 5,10,20 cases and fixed $R_\mathrm{sub}=4$ (which in reality also depends on $\xi$). }
  }
\end{figure}

\clearpage

\begin{figure}[h]
\center{
 \vspace{-0.5cm}
\includegraphics[bb=0 0 769 629,angle=0,width=0.6\textwidth,keepaspectratio]{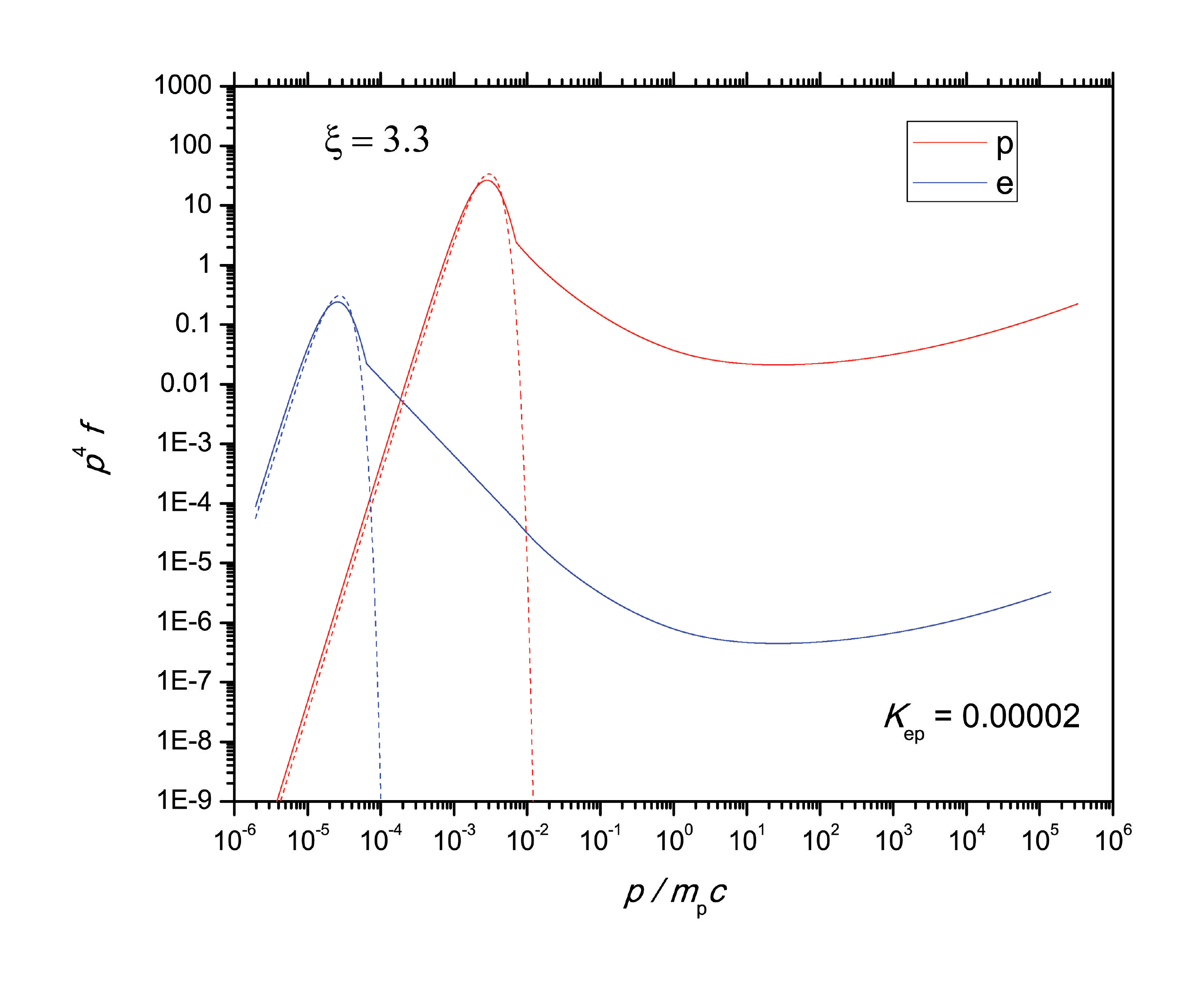}
 \vspace{-0.5cm}
\includegraphics[bb=0 0 768 626,angle=0,width=0.6\textwidth,keepaspectratio]{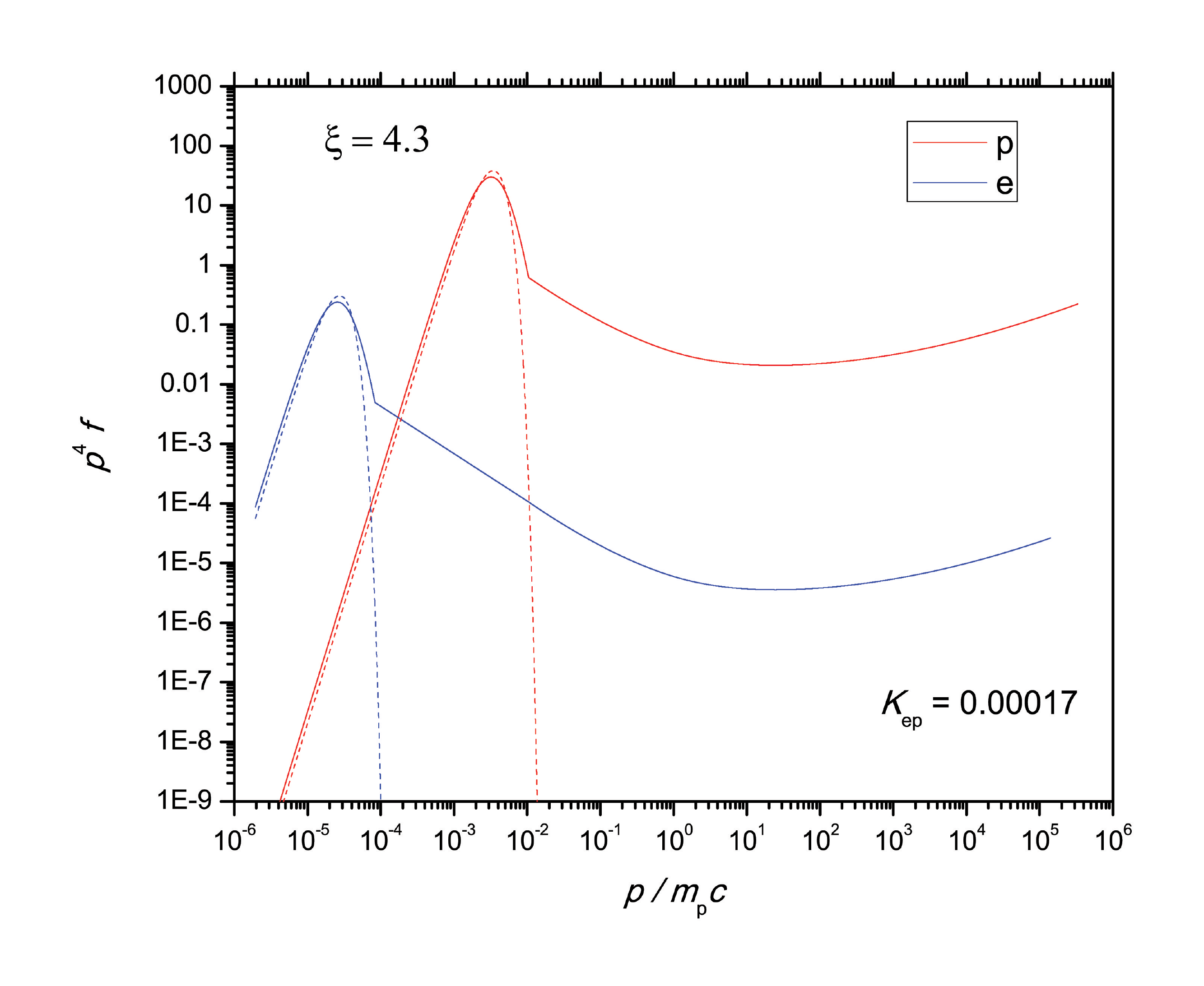}
 \vspace{-0.5cm}
\includegraphics[bb=0 0 768 626,angle=0,width=0.6\textwidth,keepaspectratio]{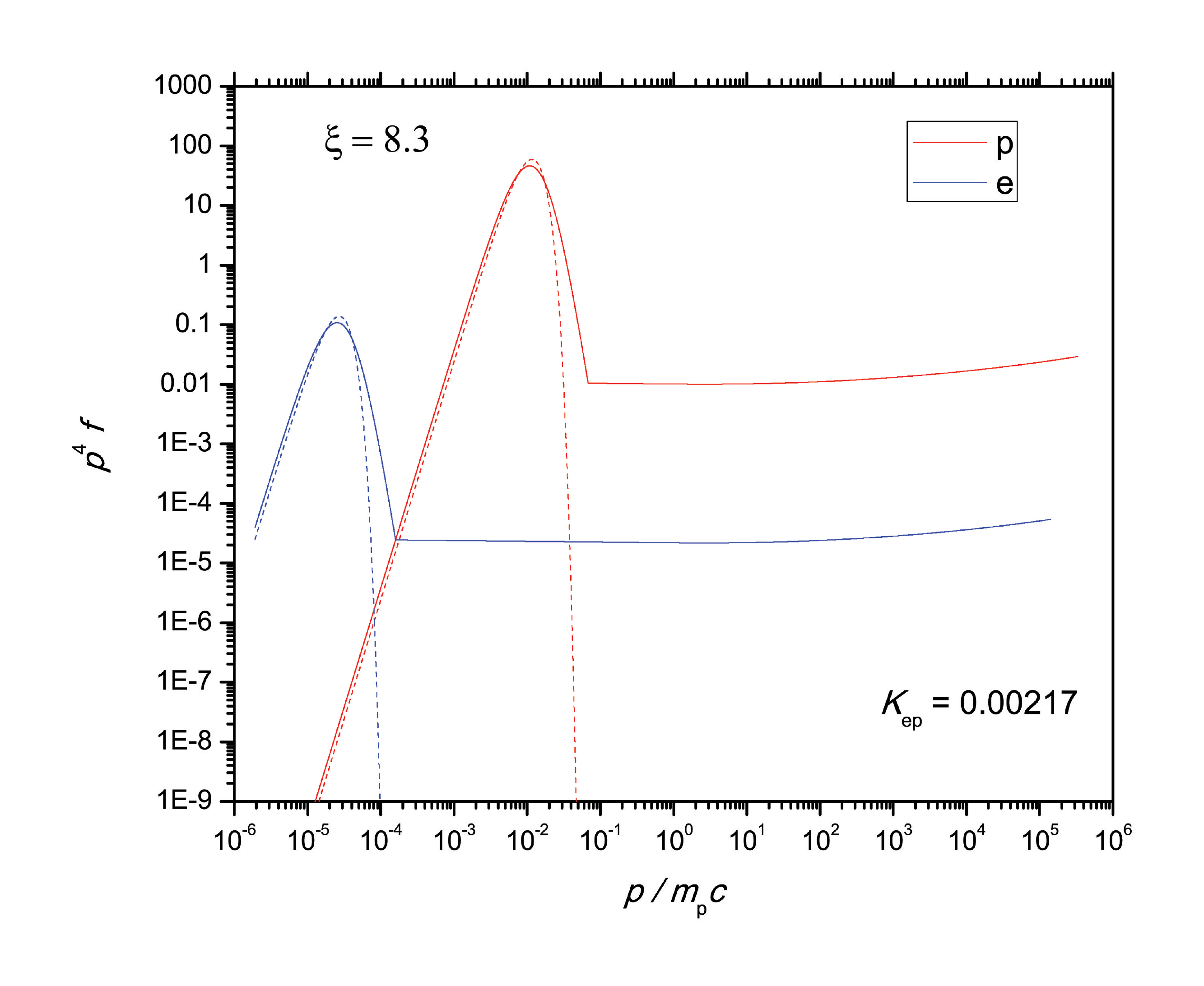}
  \vspace{-0.1cm}
  \caption{From top to bottom, proton and electron spectra for injection parameters $\xi$ = 3.3, 4.3 and  8.3, all for $\kappa = 5$. Solid line shows quasi-thermal $\kappa$-distribution and non-thermal distributions that join at $p_\mathrm{inj}$. Maxwellian with the same downstream temperature is shown with dashed line.  }
  }
\end{figure}

\clearpage

\noindent In current models \cite{Arbutina2022}, we match power-law to $\kappa$-distribution ($p_\kappa ^2 = \frac{\kappa -3/2}{\kappa} p_{\mathrm{th}}^2$) so
\begin{eqnarray}
\frac{\eta}{1-\eta } &=& \frac{4}{3\sqrt{\pi}} (R_{\mathrm{\mathrm{sub}}}-1) \frac{\Gamma(\kappa +1)}{(\kappa - \frac{3}{2})^{3/2}\Gamma(\kappa -\frac{1}{2})} \frac{\xi ^3}{\Big[ 1+ \frac{\xi^2}{\kappa -\frac{3}{2}} \Big]^{\kappa +1}} \cdot \nonumber \\
&\cdot& \Bigg[ 1-\frac{4}{\sqrt{\pi}} \frac{\Gamma(\kappa +1)}{(\kappa - \frac{3}{2})^{1/2-\kappa} (2\kappa -1) \Gamma(\kappa -\frac{1}{2})} \xi ^{1-2\kappa}\Bigg],
\end{eqnarray}
i.e.
\begin{equation}
{\eta} \approx \frac{4}{3\sqrt{\pi}} (R_{\mathrm{\mathrm{sub}}}-1) \frac{\Gamma(\kappa +1)}{(\kappa - \frac{3}{2})^{3/2}\Gamma(\kappa -\frac{1}{2})} \frac{\xi ^3}{\Big[ 1+ \frac{\xi^2}{\kappa -\frac{3}{2}} \Big]^{\kappa +1}} .
\end{equation}
The last equation generally gives higher efficiency when compared to the case $\kappa \to \infty$ (see Fig. 4).

In Fig. 5 we give the results for three cases: $\xi$ = 3.3, and $\xi$ = 4.3, as in \cite{Caprioli2010, AZ2021a}, and additionally $\xi$ = 8.3 (all for $\kappa = 5$). In all cases the other parameters are the same: shock velocity $V$ = 5000 km/s, ambient density $n_\mathrm{H} \sim$ 0.1 cm$^{-3}$, temperature $T_0$ = 100000 K, magnetic field $B_0$ = 5.3775 $\mu$Ga, Mach and Alfven-Mach numbers $M_0 = M_A = 135$, and Alfven-heating parameter $\zeta$ = 0.5. The plots show quasi-thermal $\kappa$-distribution and non-thermal distributions that join at $p_\mathrm{inj}$ (solid line) and Maxwellian with the same downstream temperature (dashed line). For the case $\xi$= 3.3, the subshock and total compressions are $R_\mathrm{sub}$ = 2.31, $R_\mathrm{tot}$ = 10.79, with acceleration efficiency $\eta$ = 0.012 and electron-to-proton ratio at high energies $K_\mathrm{ep}$ = 0.00002; while for $\xi$ = 4.3, $R_\mathrm{sub}$ = 2.67, $R_\mathrm{tot}$ = 10.75, $\eta$ = 0.0026 and $K_\mathrm{ep}$ = 0.00017. For the case $\xi$ = 8.3, the subshock and total compressions are $R_\mathrm{sub}$ = 3.96, $R_\mathrm{tot}$ = 4.91, $\eta$ = 0.00003 and $K_\mathrm{ep}$ = 0.00217.

We see that, unlike the Maxwellian-match situation where realistic $\xi \sim  3.5-4$, to reach the test particle case here, $\xi$  must be much larger and the first two cases still show characteristics of strongly modified shocks. In principle, one could introduce a modified injection parameter $\xi ' = F(\kappa)\xi$, where $F(\kappa)$ is a function chosen in such a way that values of $\xi '$ correspond to the values of original Blasi's injection parameter more, but the acceleration efficiency would still depend on two parameters $\xi '$ and index $\kappa$.

\section{Instead of conclusions}

In this contribution we gave a very brief overview of the test particle and non-linear DSA theory and the resulting non-thermal, but also supra-thermal particle spectra.  We suggest that the low-energy spectra at the shock, i.e. immediately downstream, can be represented by a single quasi-thermal distribution - the  $\kappa$-distribution that is commonly observed in out-of-equilibrium space plasmas. This may be extremely important in many "practical" applications of CR astrophysics, e.g. in modelling of the gamma or synchrotron radio emission of astrophysical sources, such as SNRs. Farther downstream, index $\kappa$ increases and the low-energy spectrum tends to Maxwell distribution \cite{AZ2021b}. On the other hand, \cite{Caprioli2020} found evidence of shock modification in hybrid simulations due to the presence of the so-called postcursor in the downstream, which has important consequences for non-thermal spectrum that need to be investigated.

The models presented can and should be further improved, e.g. in the case of electrons by including synchrotron losses \cite{Blasi2010, Diesing2019}. However, it would be more crucial to address the question of free parameters, such as $\zeta$ and $\eta$. Regarding the former, one should consider not only resonant \cite{Caprioli2009}, but also non-resonant magnetic field amplification \cite{Bell2004}, whereas regarding the latter we need to find out: a) how injection efficiencies of both protons and electrons, b) downstream electron-to-proton temperature ratio, and c) CR electron-to-proton ratio at high energies, all change with shock velocity i.e. Mach’s number. The answers to these questions can hopefully be provided by PIC simulations.

 \vspace{0.5cm}

\noindent {\bf Acknowledgements.}
The author acknowledges funding provided by the Ministry of Science, Technological Development and Innovation of the Republic of Serbia
through the contract No. 451-03-47/2023-01/200104, and support through the joint project of the Serbian Academy of Sciences and Arts and Bulgarian Academy of Sciences on the detection of Galactic and extragalactic SNRs and HII regions.

\end{document}